\documentclass[runningheads]{llncs}

\usepackage[T1]{fontenc}

\usepackage{amsmath,amssymb,amsfonts}

\usepackage{graphicx}
\usepackage[justification=centering]{subcaption}

\usepackage{booktabs}
\usepackage{array}
\usepackage{tabularx}
\usepackage{multirow}
\usepackage{makecell}
\usepackage{colortbl}

\usepackage{url}
\usepackage{csquotes}
\usepackage{comment}
\usepackage{float}
\usepackage{textcomp}

\usepackage{xcolor}

\usepackage{tikz}
\usetikzlibrary{
    arrows.meta,
    positioning,
    calc,
    shapes.geometric,
    fit
}

\usepackage{pgfplots}
\pgfplotsset{compat=1.18}
\usepgfplotslibrary{groupplots}

\usepackage[most]{tcolorbox}
\tcbuselibrary{
    theorems,
    skins,
    breakable
}

\usepackage{mdframed}
\usepackage{fancybox}

\usepackage{fontawesome5}

\usepackage{listings}

\lstdefinestyle{ide}{
    language=Python,
    backgroundcolor=\color{gray!5},
    basicstyle=\ttfamily\footnotesize,
    breaklines=true,
    showstringspaces=false,
    commentstyle=\color{green!50!black},
    stringstyle=\color{green!50!black},
    keywordstyle=\color{black}\bfseries,
    numbers=none,
    frame=single,
    rulecolor=\color{gray!30},
    frameround=ffff,
    captionpos=t
}

\providecommand{\Description}[1]{}

\definecolor{ChatColor}{RGB}{66,133,244}
\definecolor{DeepColor}{RGB}{178,34,34}

\colorlet{DeepBase}{cyan!80!black}
\colorlet{DeepFill}{cyan!80!black!28}

\colorlet{ChatBase}{orange!90!black}
\colorlet{ChatFill}{orange!90!black!28}

\definecolor{nodeFill}{RGB}{234,242,252}
\colorlet{nodeStroke}{blue!60!black}

\newtcolorbox{promptcard}[2][]{%
    enhanced,
    colback=blue!3,
    colframe=blue!60!black,
    arc=2pt,
    boxrule=0.6pt,
    left=2mm,
    right=2mm,
    top=1mm,
    bottom=1mm,
    title={#2},
    fonttitle=\bfseries,
    attach boxed title to top center={yshift=-1mm},
    boxed title style={
        colback=blue!60!black,
        colframe=blue!60!black,
        sharp corners,
        boxrule=0pt,
        top=1mm,
        bottom=1mm,
        left=1.5mm,
        right=1.5mm,
        fontupper=\color{white}
    },
    #1
}

\newtcbtheorem{user}
{\faWeixin{} \footnotesize Prompt}
{
    colback=blue!5,
    colframe=blue!35!black,
    fonttitle=\footnotesize\bfseries,
    fontupper=\footnotesize
}
{th}

\newtcolorbox{colora}{
    enhanced,
    boxrule=0pt,
    frame hidden,
    borderline west={2pt}{0pt}{gray!50!black},
    colback=gray!5!white,
    sharp corners
}

\begin{document}

\title{On the Impact of Requirement Smells in LLM-Based Code Generation}

\titlerunning{Impact of Requirement Smells in LLM-Based Code Generation}

\author{
Hugo Villamizar\inst{1}\orcidID{0000-0003-4142-6967}
\and
Jannik Fischbach\inst{2,1}\orcidID{0000-0002-4361-6118}
\and
Mert Şahin\inst{3}
\and
Julian Frattini\inst{4}\orcidID{0000-0003-3995-6125}
\and
Alessio Ferrari\inst{5}\orcidID{0000-0002-0636-5663}
\and
Alexander Korn\inst{6}\orcidID{0009-0002-6258-6791}
\and
Andreas Vogelsang\inst{6}\orcidID{0000-0003-1041-0815}
\and
Daniel Mendez\inst{7,1}\orcidID{0000-0003-0619-6027}
}

\authorrunning{H. Villamizar et al.}

\institute{
fortiss GmbH, Munich, Germany\\
\email{guarinvillamizar@fortiss.org}
\and
Netlight Consulting, Munich, Germany\\
\email{jannik.fischbach@netlight.com}
\and
Technical University of Munich, Munich, Germany\\
\email{mert.sahin@tum.de}
\and
Chalmers University of Technology, Gothenburg, Sweden\\
\email{julian.frattini@chalmers.se}
\and
Trinity College Dublin, Dublin, Ireland\\
\email{aferrari@tcd.ie}
\and
University of Duisburg-Essen, Essen, Germany\\
\email{\{alexander.korn,andreas.vogelsang\}@uni-due.de}
\and
Blekinge Institute of Technology, Karlskrona, Sweden\\
\email{daniel.mendez@bth.se}
}

\maketitle

\begin{abstract}
Software requirements are typically incorporated into prompts used in LLM-assisted software development. Recent work has shown that requirement smells can affect automated traceability between requirements and code, but empirical evidence on their effects in code generation remains limited. To address this gap, we build upon a prior study on automated traceability by reusing its dataset and requirement smell taxonomy, while extending it to evaluate the functional correctness of LLM-generated code. Using a benchmark consisting of requirements and corresponding system tests for four applications, we progressively introduced semantic, syntactic, and lexical smells into otherwise clear requirements and analyzed their influence on generated implementations. Our results suggest that increasing \textit{smell density} was generally associated with lower test-suite-based functional correctness, although non-smelly requirements could still produce faulty code. We also found that different smell categories had similar effects. These findings provide additional empirical evidence of the importance of requirement quality in LLM-assisted code generation, while showing that high-quality requirements alone do not guarantee correctness, as these depends on several factors, including the LLM. Compared with previous work, our results suggest that the impact of requirement smells depends on the software engineering task: whereas their effects on traceability were modest, code generation appears more sensitive. Overall, this work motivates further investigation into task-dependent quality effects in LLM-assisted software engineering.

\keywords{
Requirements quality
\and
Requirement smells
\and
Prompt quality
\and
Large language models
\and
Code generation
}
\end{abstract}

\section{Introduction}
\label{sec:introduction}

Large language models (LLMs) are widely used to support tasks across the entire software development lifecycle~\cite{stackoverflow2025survey}. By learning from massive corpora including natural language and code, these models can, among other tasks, generate functional implementations from high-level descriptions, accelerating development and lowering the barrier to entry for non-experts. However, their performance is known to depend on the quality of the prompts they receive, highlighting the role of software engineering (SE) practices in shaping LLM-generated outputs~\cite{hassan2024rethinking,nahar2025beyond}.

In many AI-assisted SE workflows, software requirements are used as structured prompts that guide LLM behavior. As a result, characteristics traditionally studied in requirements engineering (RE), such as ambiguity, inconsistency, and requirement smells, may directly influence the quality of LLM-generated artifacts. Recent work has shown that requirement smells can affect LLM-supported automated traceability between requirements and code, with effects varying across tracing tasks and smell categories~\cite{vogelsang2025impact}. More broadly, studies on prompt quality have shown that its specificity, structure, and requirement articulation influence the correctness of LLM-generated code~\cite{ma2025should,murr2023testingllmscodegeneration,yang2025prompts,dellaportapromptqualitygithub}. However, despite these advances, empirical evidence on how requirement smells affect LLM-based code generation remains limited.


To address this gap, we present a controlled experiment that builds upon and extends the work of Vogelsang \textit{et al.}~\cite{vogelsang2025impact}, which focuses on the impact of requirements smells on requirements traceability. Specifically, we reuse the dataset, requirement smell taxonomy, and smell injection strategy from the original study while extending the evaluation from automated traceability to LLM-based code generation. Using a benchmark consisting of natural language requirements and corresponding system tests for four applications, we progressively introduce semantic, syntactic, and lexical requirement smells into otherwise clear requirements and evaluate their impact on the test-suite-based functional correctness of LLM-generated code produced by two state-of-the-art LLMs. 

Our results suggest that increasing smell density was generally associated with lower test-suite-based functional correctness, while the effects of individual smell categories varied across LLMs and applications. Our results further suggest that the effects of requirement smells are context-dependent, emerging from the interaction between the application, the LLM, and the requirement formulation rather than from the smell category alone. These findings provide additional exploratory empirical evidence that requirements quality matters for code generation, thus highlighting RE as an important factor in shaping the reliability of LLM-assisted software development. 



\noindent
All material, including the source code, experimental data, prompts, and replication package, is publicly available.\footnote{\url{https://doi.org/10.5281/zenodo.17441075}}
\section{Background and Related Work}
\label{sec:background-and-related-work}


Within RE, the notion of \textit{requirement smells} has long been used to capture potential quality problems in requirements specifications~\cite{femmer2017rapid}. Similar to code smells, requirement smells are indicators of possible defects that may negatively affect downstream development activities. Prior research has shown that requirement smells can manifest at different levels, including lexical, syntactic, and semantic issues, and may impact activities such as design, testing, maintenance, and communication~\cite{femmer2018requirements,frattini2023requirements}. 



At the same time, recent research on LLMs has increasingly emphasized the importance of prompt quality for achieving reliable and reproducible outputs~\cite{sclar2024quantifying,cao2024worst}. In SE, prompts are often constructed from existing development artifacts, including software requirements, user stories, and functional specifications. Ullrich \textit{et al.}~\cite{ullrich2025requirements} observed that practitioners frequently operationalize and refine requirements into more concrete and context-rich prompts to support LLM-based code generation. 
Building on this connection between requirements and prompts, Vogelsang \textit{et al.}~\cite{vogelsang2025impact} investigated whether requirement smells affect LLM-based automated traceability between requirements and code. Their results showed mixed effects depending on the tracing task and smell category, providing initial evidence that quality issues originating from requirements may influence LLM-enabled SE tasks.

Beyond requirements specifically, several recent studies have investigated how broader prompt quality characteristics relate to LLM-generated code. Murr \textit{et al.}~\cite{murr2023testingllmscodegeneration} showed that prompt ambiguity and reduced specificity can negatively affect the accuracy of LLM-generated code, while Yang \textit{et al.}~\cite{yang2025prompts} found that underspecified prompts reduce the robustness and reproducibility of LLM outputs. Ma \textit{et al.}~\cite{ma2025should} demonstrated that improving requirement articulation can enhance prompt effectiveness in code generation settings. Della Porta \textit{et al.}~\cite{dellaportapromptqualitygithub} analyzed real-world developer interactions with ChatGPT and found that prompt readability was associated with correctness and perceived usefulness, while structural characteristics were associated with the conceptual consistency of generated code. Together, these studies provide growing evidence that different dimensions of prompt quality are associated with the quality of LLM-generated code. 

Overall, existing studies provide growing evidence that prompt quality influences LLM-enabled SE tasks. However, more empirical evidence is needed to strengthen this understanding. We contribute to this body of evidence by examining requirement smells as a specific dimension of prompt quality and investigating how their presence, density, and category relate to the functional correctness of LLM-generated code.

\section{Study Design}
\label{sec:study-design}

This study follows established guidelines for controlled experiments in SE~\cite{wohlin2012experimentation} to investigate how the quality of software requirements used in prompts affects LLM-based code generation.

\subsection{Research Goal}

Following the Goal--Question--Metric (GQM) approach~\cite{caldiera1994goal}, the goal of this study is:

\begin{quote}
\textit{Analyze LLM-based code generation for the purpose of understanding the impact of requirement smells on the correctness of generated implementations with respect to test-suite-based functional correctness from the viewpoint of SE researchers in the context of controlled experiments using natural language software requirements.}
\end{quote}

To achieve this goal, we formulate the following three complementary research questions.

\subsection{Research Questions}

\begin{itemize}
\setlength\itemsep{0.8em}
\renewcommand\labelitemi{}

\item \textbf{RQ1. How well can LLMs generate functionally correct code when given non-smelly requirements?} This question establishes the baseline capability of the evaluated LLMs under ideal input conditions. It allows us to assess how reliably the LLMs can implement software requirements when no defects are present in the prompt.

\item \textbf{RQ2. How does the presence of requirement smells in prompts affect the functional correctness of LLM-generated code?} 
This question examines whether and how the presence of requirement smells reduces correctness. By injecting requirement smells into prompts, we assess whether degradation in correctness occurs and whether it follows a measurable pattern.

\item \textbf{RQ3. How do different categories of requirement smells impact the functional correctness of LLM-generated code?} 
This question investigates whether some smell categories (\textit{e.g.}, semantic vs.\ syntactic vs.\ lexical) have stronger or more consistent effects on correctness than others, or whether their influence varies across tasks.

\end{itemize}

\subsection{Experimental Factors and Variables}

Table~\ref{tab:replication} summarizes the conceptual replication strategy adopted in this study, highlighting the components reused from the original study by Vogelsang \textit{et al.}~\cite{vogelsang2025impact} and the elements introduced to support the evaluation of LLM-based code generation. Table~\ref{tab:exp-variables} summarizes the experimental variables considered throughout the study. 

\begin{table}[t]
\centering
\scriptsize
\caption{Conceptual replication strategy.}
\label{tab:replication}

\setlength{\tabcolsep}{7pt}

\begin{tabular}{p{2.2cm}p{3.8cm}p{4cm}}
\toprule
\textbf{Component} & \textbf{Original Study} & \textbf{This Study} \\
\midrule
Study objects & Five game applications & Four reused \\
Requirements & 94 functional reqs. & 74 reused \\
Smell taxonomy & Frattini \textit{et al.} taxonomy~\cite{frattini2022live} & Reused \\
Smell injection & Manual smell injection & Reused \\
Prompt objective & Trace link generation & Java code generation \\
Evaluation & BTA, LOC precision/recall/F1 & Requirement-level test pass rates \\
Java skeleton & -- & Added \\
Unit tests & -- & Added \\
\bottomrule
\end{tabular}
\end{table}

\begin{table}[t]
\centering
\scriptsize
\caption{Experimental variables.}
\label{tab:exp-variables}
\begin{tabular}{@{}p{1.8cm}p{3cm}p{5.5cm}@{}}
\toprule
\textbf{Type} & \textbf{Variable} & \textbf{Values} \\ \midrule
Independent & Smell presence & Yes / No \\
& Smell density & Sets 1–5 (None / Low–Medium / High) \\
& Smell category & Semantic / Syntactic / Lexical \\
& LLM & GPT-4o / DeepSeek-V3 \\ \midrule
Dependent & Functional correctness & Unit test pass rate (Avg / Best) \\
& Stability & Variance, standard deviation (SD) \\ \midrule
Controlled & Prompt structure & Fixed template across all games \\
& LLM configuration & Fixed parameters (\textit{e.g.}, temperature = 0)\\ 
\midrule
Context & Game type & Dice / Arkanoid / Snake / Scopa \\
& Skeleton code & Fixed per game, variable between games \\
\bottomrule
\end{tabular}
\end{table}

\subsection{Game Applications and Objects of Study}

In this experiment, four game applications serve as the domain for evaluating both LLMs: \textit{Dice}, a turn-based dice game involving player turns, dice rolls, and score calculation; \textit{Arkanoid}, brick-breaking arcade game involving paddle movement, ball interactions, and collision handling; \textit{Snake}, in which a player-controlled snake grows by collecting items while avoiding collisions; and \textit{Scopa}, a traditional Italian card game involving card matching, capturing, and scoring. Their use provides a controlled experimental setting that allows us to isolate the variables of interest and helps reduce additional sources of variability, such as heterogeneous application complexity, domain-specific knowledge, and tacit requirements.

Each game provides a set of functional requirements that define the expected system behavior and form the basis for code generation. The objects of study are these requirements, which serve as the primary units of manipulation in the experiment, including both non-smelly baseline specifications and smell-injected variants. Table~\ref{tab:benchmark-overview} summarizes the dataset and the two smell configurations used in the experiment, which are detailed in Section~\ref{sec:experimental-procedure}.

\begin{table}[t]
\centering
\scriptsize
\caption{Benchmark overview and smell configurations used for RQ2 and RQ3. For RQ2, the five values correspond to Sets 1--5, respectively.}
\label{tab:benchmark-overview}

\setlength{\tabcolsep}{7pt}

\begin{tabular}{@{}lccccc@{}}
\toprule
\textbf{Game} &
\textbf{Total Reqs.} &
\textbf{RQ2: No. of smells per set} &
\multicolumn{3}{c}{\textbf{Smells per RQ3 category set}} \\
\cmidrule(lr){4-6}
&
&
&
\textbf{Semantic} &
\textbf{Syntactic} &
\textbf{Lexical} \\
\midrule
\textit{Dice}     & 25 & 0, 5, 9, 13, 17 & 4 & 4 & 4 \\
\textit{Arkanoid} & 19 & 0, 4, 7, 10, 13 & 4 & 4 & 4 \\
\textit{Snake}    & 14 & 0, 3, 5, 7, 9   & 2 & 2 & 2 \\
\textit{Scopa}    & 16 & 0, 3, 6, 9, 12  & 3 & 3 & 3 \\
\bottomrule
\end{tabular}
\end{table}


\begin{table}[t]
\centering
\scriptsize

\renewcommand{\arraystretch}{2}
\setlength{\tabcolsep}{4pt}

\caption{Examples of non-smelly and smelly requirements for \textit{Dice}.}
\label{tab:dice-reqs-comparison}

\begin{tabular}{@{}p{0.4cm}p{3.8cm}p{3.8cm}p{1.7cm}p{1.5cm}@{}}
\toprule

\textbf{ID} &
\textbf{Non-smelly requirement} &
\textbf{Smelly requirement} &
\textbf{Type} &
\textbf{Category} \\

\midrule

9 &
If and only if the dice count is 1, the player must decide if he wants to re-roll the dice instead.
&
If and only if the dice count is 1, it must be decided if the player re-rolls the dice instead.
&
passive\_voice &
syntactic \\

12 &
If a player's points get above 11, his point color shall turn purple.
&
If a player's points are sufficient to win, his point color shall turn purple.
&
weak\_verbs &
lexical \\

14 &
If the dice count is 2, it shall be tripled. If the dice count is even but not 2, it shall be halved.
&
If the dice count is 2, it shall be tripled. If the dice count is even, it shall be halved.
&
negative &
syntactic \\

24 &
The player with the most points after the game has ended will be the winner.
&
After the game has ended, one player shall be the winner.
&
ambiguities &
semantic \\

\bottomrule
\end{tabular}

\end{table}

Smells were organized into three categories: \textit{semantic smells}, which affect the meaning or interpretation of a requirement, for example by introducing ambiguity or context-dependent interpretations; \textit{syntactic smells}, which concern how a requirement is grammatically or structurally formulated; and \textit{lexical smells}, which arise from the use of individual words or expressions that may reduce precision or clarity. These categories were selected because they capture different abstraction levels of defects commonly discussed in RE~\cite{frattini2022live,femmer2014rapid}. Table~\ref{tab:dice-reqs-comparison} illustrates examples of non-smelly and smell-injected requirements for the \textit{Dice} game used to operationalize \textit{RQ2} and \textit{RQ3}. The examples illustrate different ways in which requirement quality may be degraded. For instance, the passive-voice example (\textit{ID~9}) removes the explicit actor responsible for the decision, making the requirement less direct. Likewise, the weak-verb example (\textit{ID~12}) replaces a precise numerical threshold with the vague expression ``sufficient to win'', increasing interpretive flexibility while preserving grammatical correctness.

    


\subsection{Experimental Procedure}
\label{sec:experimental-procedure}

\textbf{Smell injection.} For RQ2, we constructed five requirement sets per game application with an increasing number of smells, ranging from a clean baseline (Set~1) to the maximum number of smelly requirements available for each game (Sets~2–5). For RQ3, three separate category-specific sets were constructed using an equal number of semantic, syntactic, and lexical smells within each game. The number of smells per category was limited by the least represented category available for that game. 

Smell-injected variants were manually designed based on an established requirement smell taxonomy~\cite{frattini2022live}. From the broader set of quality defects represented in the taxonomy, we selected smells that could be realistically introduced while preserving the original requirement intent. The selection and resulting variants were further informed by prior empirical research on requirements quality~\cite{montgomery2022empirical} and the authors' research and industrial experience. Each modified requirement preserved the original intent while introducing only the intended smell. The resulting variants were independently reviewed by another author to verify that (i) the injected smell corresponded to the intended category, (ii) no unintended defects were introduced, and (iii) the resulting requirement remained realistic. Disagreements were resolved through discussion until consensus was reached.

\textbf{Skeleton Definition.}
Each game was associated with a predefined Java skeleton containing classes, attributes, and method signatures linked to specific requirements. Each requirement was mapped to exactly one predefined method in the skeleton, enabling the corresponding behavior to be evaluated independently through requirement-level unit tests. A method could nevertheless be linked to multiple requirements when those requirements specified distinct conditions or aspects of the same functionality.

\textbf{Prompt Construction.}
Figure~\ref{fig:prompt-structure} illustrates the structure of the prompt provided to the LLM. For each game application, we constructed a prompt containing four components: (i) a task description, (ii) implementation constraints, (iii) the predefined Java skeleton, and (iv) the complete set of functional requirements corresponding to one experimental condition. The first three components remained identical across all experiments, while only the functional requirements were modified through the controlled injection of semantic, syntactic, and lexical requirement smells.

\begin{figure}[t]
\Description{Schematic representation of the prompt provided to the LLM. The task description, implementation constraints, and Java skeleton were held constant, while the functional requirements were varied through requirement smell injection.}
\centering

\begin{tikzpicture}[
  font=\scriptsize,
  component/.style={
    draw=black!55,
    rounded corners=2pt,
    align=left,
    text width=0.90\columnwidth,
    inner xsep=6pt,
    inner ysep=4pt
  },
  fixed/.style={
    component,
    fill=black!2
  },
  varied/.style={
    component,
    draw=black,
    line width=0.9pt,
    fill=black!7
  },
  outer/.style={
    draw=black!75,
    rounded corners=3pt,
    line width=0.8pt,
    inner sep=5pt
  }
]

\node[fixed] (task) {
  \textbf{1. Task description}
  \hfill\textit{(held constant)}\\[-1pt]
  Develop a Java 8 application for an Arkanoid game in which the player
  controls a paddle to hit a ball and destroy bricks.
};

\node[fixed, below=2.5pt of task] (constraints) {
  \textbf{2. Implementation constraints}
  \hfill\textit{(held constant)}\\[-1pt]
  Implement all classes and methods according to the provided skeleton;
  return the complete code in one file; and follow the specified visibility
  and structural constraints.
};

\node[fixed, below=2.5pt of constraints] (skeleton) {
  \textbf{3. Predefined Java skeleton}
  \hfill\textit{(held constant)}\\[-1pt]
  \texttt{class Ball}\\
  \hspace*{1em}\texttt{int x, y, diameter, xSpeed, ySpeed;}\\
  \hspace*{1em}\texttt{Ball(int x, int y);}\\
  \hspace*{1em}\texttt{void move();}
  \hfill\textit{Implements R4, R12, R13}\\[-1pt]
  \hspace*{1em}\texttt{\ldots}
};

\node[varied, below=2.5pt of skeleton] (requirements) {
  \textbf{4. Functional requirements}
  \hfill\textit{(experimentally varied)}\\[-1pt]
  \textbf{R1.} The game board shall have the specified dimensions.\\
  \textbf{R2.} The paddle shall move horizontally within the board limits.\\
  \textbf{R3.} The ball shall change direction when it hits the paddle.\\
  \hspace*{1em}\textit{\ldots}\\[-1pt]
  \textit{Semantic, syntactic, or lexical smells were injected into this component.}
};

\node[
  outer,
  fit=(task)(constraints)(skeleton)(requirements)
] {};

\end{tikzpicture}

\caption{Excerpt of the prompt structure used for \textit{Arkanoid}.}
\label{fig:prompt-structure}
\end{figure}

\textbf{LLM Configuration.} We evaluated GPT-4o from OpenAI and DeepSeek-V3. They were selected because they represent two state-of-the-art LLM families that were publicly available during the study: one proprietary (GPT-4o) and one open-weight (DeepSeek-V3), both exhibiting competitive code-generation capabilities~\cite{ahmedbenchmarking2025}. Both were accessed via their APIs using comparable configurations to ensure comparability. We set the temperature parameter to $0$ to reduce stochastic variability. No explicit maximum token limit was set, models generated outputs up to their allowed context window. While GPT-4o is closed source and its internal caching mechanisms are opaque, DeepSeek-V3 uses a disk-based context caching mechanism. Each LLM was prompted under three conditions: (\emph{i}) a non-smelly baseline, (\emph{ii}) five increasing levels of smell density, and (\emph{iii}) individual smell categories. For each configuration, five outputs were generated to capture residual variation, yielding deterministic but replicable results.

\textbf{Unit Testing Strategy.} To ensure a consistent and objective evaluation of generated implementations, we adopted a requirement-level unit testing strategy. Each natural language requirement was mapped to one or more unit tests implemented in Java using \texttt{JUnit}, while \texttt{Mockito} was used whenever controlled execution of external interactions or edge cases was required. The complete test suite was developed specifically for this study and peer-reviewed by two authors to verify that each test unambiguously captured the intended requirement behavior. Table~\ref{tab:test-strategy} presents representative examples from the \textit{Dice} game, illustrating different testing strategies, including direct state verification, mocked executions, and validation of game termination logic. Since each prompt produced a complete program, the entire test suite was executed for every generated implementation, and the number of passed requirement-level tests was used to assess test-suite-based functional correctness.

\begin{table}[t]
\centering
\scriptsize
\renewcommand{\arraystretch}{1.25}
\setlength{\tabcolsep}{5pt}

\caption{Excerpt of requirement-level test strategy for \textit{Dice}.}
\label{tab:test-strategy}

\begin{tabular}{@{}cp{3.5cm}p{7.4cm}@{}}
\toprule
\textbf{ID} &
\textbf{Function Tested} &
\textbf{Testing Strategy} \\
\midrule

1 &
\texttt{initializeGame()} &
Simulates user input for player registration and verifies that the correct number of players is initialized. \\

2 &
\texttt{testPlayerInitialPoints()} &
Verifies that all players start the game with zero points immediately after initialization by checking the object state. \\

6 &
\texttt{testPlayerTurnConditions6()} &
Mocks dice rolls and verifies that the printed output reflects the expected dice value and updated score. \\

8 &
\texttt{testGameEndConditionsReq8()} &
Manipulates player scores to trigger the end-of-game condition and verifies the correct game termination behavior. \\

12 &
\texttt{testAutomaticColorChange12()} &
Adjusts player points to verify that the player's color changes to purple when the specified threshold is reached. \\

18 &
\texttt{testDiceMultiply18()} &
Mocks dice rolls to force a predefined edge case and verifies the multiplication logic and point assignment. \\

25 &
\texttt{testNewGamePreparation25()} &
Simulates the end of a match and verifies winner announcement and preparation of the next game. \\

\bottomrule
\end{tabular}
\end{table}

\textbf{Data Recording.} For each generated code artifact, the outcome of every requirement-level unit test was automatically recorded, capturing whether the implementation passed or failed each test. These results were stored together with metadata identifying the LLM, game application, smell condition (baseline, density level, smell category), and run index. The collected data were then aggregated by (\emph{i}) requirement to analyze how individual requirements were affected by smells and (\emph{ii}) requirement set to compute average and best-case performance metrics per configuration. This structure enabled both fine-grained and aggregate analyses of LLM correctness across different experimental conditions.

\subsection{Evaluation Metrics and Analysis}
\label{subsec:evaluation-metrics}

The evaluation focuses on two key constructs: (\emph{i}) \emph{requirement quality}, operationalized through the presence or absence of requirement smells in the input prompt, and (\emph{ii}) \emph{LLM output quality}, assessed through the test-suite-based functional correctness of the generated code. Requirement quality was operationalized through categorical experimental conditions, including non-smelly requirements and smell-injected variants containing lexical, syntactic, or semantic defects. 

A requirement was considered correctly implemented when the corresponding unit test passes. For each condition, game application, and LLM, test outcomes were aggregated into two complementary metrics: \textit{average performance}, defined as the mean percentage of passed tests across the five runs, and \textit{best performance}, defined as the highest percentage of passed tests obtained in a single run.

For \textit{RQ1}, we computed pass rate, variance, and standard deviation across runs using non-smelly requirements. Pass rate reflects the proportion of correctly implemented requirements, while variance and standard deviation capture the stability of LLM outputs under repeated executions. For \textit{RQ2}, we analyzed the relationship between smell density and test-suite-based functional correctness using descriptive statistics and Spearman’s rank-order correlation ($\rho$). Statistical significance was assessed using a two-sided test with $\alpha = 0.05$ and Bonferroni correction across $K = 15$ correlations ($\alpha^* = 0.0033$). For \textit{RQ3}, we compared smelly and non-smelly requirements using chi-square tests of independence and computed Cramér's V to assess effect size~\cite{cohen2013statistical}. Bonferroni correction was applied separately to the aggregate and application-level analyses.

\section{Results}
\label{sec:results}

\subsection{LLM Functional Correctness with Non-Smelly Requirements}

Figure~\ref{fig:rq1} presents the performance of GPT-4o and DeepSeek-V3 when provided with non-smelly requirements. Both LLMs achieved their highest pass rates for \textit{Arkanoid}. DeepSeek-V3 passed all 19 requirements in three of the five runs, while GPT-4o passed up to 18. For \textit{Dice}, the number of passed requirement-level tests ranged from 17 to 23 across runs and models, although neither LLM achieved complete correctness.

For \textit{Snake}, each LLM produced the same pass rate across all five runs, resulting in zero variance. Nevertheless, both LLMs consistently failed to implement some requirements, including behaviors related to timed events. \textit{Scopa} exhibited greater variation across runs, particularly for GPT-4o, whose average pass rate was 72.5\% with a standard deviation of 15.0 percentage points. DeepSeek-V3 achieved a higher average pass rate of 86.3\% with less variation.

Requirement-level inspection showed that some recurring failures involved requirements mapped to shared skeleton functions. In \textit{Dice}, for example, requirements describing related scoring and end-game conditions were sometimes only partially integrated into the same function. Similar patterns occurred in \textit{Scopa} for requirements involving multiple rounds. These observations indicate possible interactions between requirement dependencies, their ordering in the prompt, and the structure of the provided skeleton. Because these factors were not independently manipulated, they are treated as potential confounding factors rather than effects attributable to requirement smells.

\begin{tcolorbox}[colback=gray!5!white, colframe=gray!70!white, title=\textbf{Answer to RQ1}]
\textit{Under non-smelly conditions, both evaluated LLMs generated implementations that passed a majority of the requirement-level tests, although performance varied across game applications. DeepSeek-V3 achieved higher average pass rates in three of the four applications, while recurring failures indicate that clean requirements alone do not guarantee complete functional correctness.}
\end{tcolorbox}

These results establish the non-smelly baseline against which the smell-density and smell-category conditions are evaluated in RQ2 and RQ3.

\subsection{Effect of Smell Density on LLM Functional Correctness}

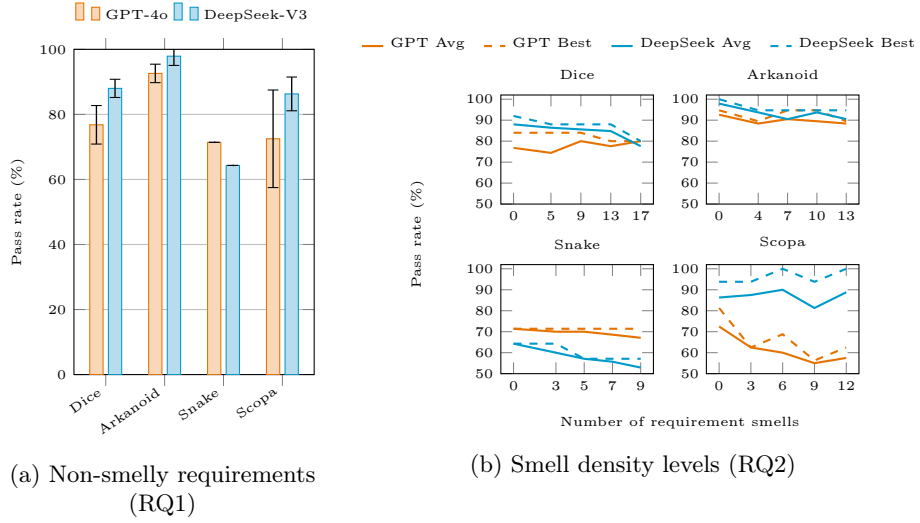
\begin{figure}[t]
\Description{Combined figure showing baseline pass rates on non-smelly requirements and pass rates across smell density levels.}
\centering
\scriptsize

\begin{subfigure}[c]{0.39\columnwidth}
\centering

\begin{tikzpicture}
\begin{axis}[
  ybar,
  ymin=0, ymax=100,
  height=5.88cm,
  width=\linewidth,
  bar width=5pt,
  symbolic x coords={Dice,Arkanoid,Snake,Scopa},
  xtick=data,
  ymajorgrids=true,
  enlarge x limits=0.18,
  ylabel={Pass rate (\%)},
  tick label style={font=\tiny},
  label style={font=\tiny},
  x tick label style={rotate=35,anchor=east},
  legend style={font=\tiny,at={(0.5,1.05)},anchor=south,legend columns=2,draw=none}
]
\addplot[
  ybar, fill=ChatFill, draw=ChatBase, mark=none,
  error bars/.cd, y dir=both, y explicit,
  error bar style={line width=0.5pt, draw=black}
]
coordinates {
 (Dice,76.8) +- (0,5.92)
 (Arkanoid,92.6) +- (0,2.84)
 (Snake,71.4) +- (0,0.00)
 (Scopa,72.5) +- (0,15.00)
};
\addplot[
  ybar, fill=DeepFill, draw=DeepBase, mark=none,
  error bars/.cd, y dir=both, y explicit,
  error bar style={line width=0.5pt, draw=black}
]
coordinates {
 (Dice,88.0) +- (0,2.80)
 (Arkanoid,97.9) +- (0,2.84)
 (Snake,64.3) +- (0,0.00)
 (Scopa,86.3) +- (0,5.19)
};
\legend{GPT-4o, DeepSeek-V3}
\end{axis}
\end{tikzpicture}

\caption{Non-smelly requirements (RQ1)}
\label{fig:rq1}
\end{subfigure}
\hfill
\begin{subfigure}[c]{0.59\columnwidth}
\centering

{\tiny
\begin{tabular}{@{}l@{\quad}l@{\quad}l@{\quad}l@{}}
  \tikz{\draw[ChatBase,thick] (0,0)--(.28,0);} GPT Avg &
  \tikz{\draw[ChatBase,thick,dashed] (0,0)--(.28,0);} GPT Best &
  \tikz{\draw[DeepBase,thick] (0,0)--(.28,0);} DeepSeek Avg &
  \tikz{\draw[DeepBase,thick,dashed] (0,0)--(.28,0);} DeepSeek Best
\end{tabular}
}

\vspace{-3pt}

\begin{tikzpicture}
\begin{groupplot}[
  group style={
    group size=2 by 2,
    horizontal sep=7mm,
    vertical sep=8mm,
  },
  width=0.5\linewidth,
  height=0.42\linewidth,
  ymin=50,
  ymax=102,
  ytick={50,60,70,80,90,100},
  title style={font=\tiny, yshift=-3pt},
  every axis/.append style={
    label style={font=\tiny},
    tick label style={font=\tiny},
  },
]

\nextgroupplot[title={Dice}, xtick={0,5,9,13,17}]
\addplot[ChatBase, thick] coordinates {(0,76.8) (5,74.4) (9,80.0) (13,77.6) (17,80.0)};
\addplot[ChatBase, thick, dashed] coordinates {(0,84.0) (5,84.0) (9,84.0) (13,80.0) (17,80.0)};
\addplot[DeepBase, thick] coordinates {(0,88.0) (5,86.4) (9,85.6) (13,84.8) (17,77.6)};
\addplot[DeepBase, thick, dashed] coordinates {(0,92.0) (5,88.0) (9,88.0) (13,88.0) (17,80.0)};

\nextgroupplot[title={Arkanoid}, xtick={0,4,7,10,13}]
\addplot[ChatBase, thick] coordinates {(0,92.6) (4,88.4) (7,90.5) (10,89.5) (13,88.4)};
\addplot[ChatBase, thick, dashed] coordinates {(0,94.7) (4,89.5) (7,94.7) (10,94.7) (13,89.5)};
\addplot[DeepBase, thick] coordinates {(0,97.9) (4,93.7) (7,90.5) (10,93.7) (13,90.5)};
\addplot[DeepBase, thick, dashed] coordinates {(0,100.0) (4,94.7) (7,94.7) (10,94.7) (13,94.7)};

\nextgroupplot[title={Snake}, xtick={0,3,5,7,9}]
\addplot[ChatBase, thick] coordinates {(0,71.4) (3,70.0) (5,70.0) (7,68.6) (9,67.1)};
\addplot[ChatBase, thick, dashed] coordinates {(0,71.4) (3,71.4) (5,71.4) (7,71.4) (9,71.4)};
\addplot[DeepBase, thick] coordinates {(0,64.3) (3,60.0) (5,57.1) (7,55.7) (9,52.9)};
\addplot[DeepBase, thick, dashed] coordinates {(0,64.3) (3,64.3) (5,57.1) (7,57.1) (9,57.1)};

\nextgroupplot[title={Scopa}, xtick={0,3,6,9,12}]
\addplot[ChatBase, thick] coordinates {(0,72.5) (3,62.5) (6,60.0) (9,55.0) (12,57.5)};
\addplot[ChatBase, thick, dashed] coordinates {(0,81.3) (3,62.5) (6,68.8) (9,56.3) (12,62.5)};
\addplot[DeepBase, thick] coordinates {(0,86.3) (3,87.5) (6,90.0) (9,81.3) (12,88.8)};
\addplot[DeepBase, thick, dashed] coordinates {(0,93.8) (3,93.8) (6,100.0) (9,93.8) (12,100.0)};

\end{groupplot}

\node[rotate=90, anchor=center, font=\tiny]
  at ($(group c1r1.west)!0.5!(group c1r2.west) + (-1.1cm, 0)$)
  {Pass rate (\%)};

\node[anchor=center, font=\tiny]
  at ($(group c1r2.south)!0.5!(group c2r2.south) + (0, -0.65cm)$)
  {Number of requirement smells};

\end{tikzpicture}

\caption{Smell density levels (RQ2)}
\label{fig:rq2}
\end{subfigure}

\caption{Pass rate on non-smelly reqs. (left) and pass rates across smell density levels (right).}
\label{fig:rq1-rq2-combined}
\end{figure}

Figure~\ref{fig:rq2} shows the average and best pass rates across the five smell-density levels. The descriptive results show decreasing average pass rates in several LLM--application configurations, although the magnitude, monotonicity, and even direction of the observed patterns vary across applications and LLMs.

For \textit{Dice}, the average pass rate of DeepSeek-V3 decreased from 88.0\% for the non-smelly requirements to 77.6\% at the highest density level. GPT-4o fluctuated between 74.4\% and 80.0\%, without a clear monotonic pattern. For \textit{Arkanoid}, both LLMs maintained comparatively high performance across all density levels: GPT-4o remained above 88\%, while DeepSeek-V3 remained above 90\%.

For \textit{Snake}, average performance decreased across the smell-density levels for both LLMs. GPT-4o declined from 71.4\% to 67.1\%, while DeepSeek-V3 declined from 64.3\% to 52.9\%. For \textit{Scopa}, GPT-4o decreased from 72.5\% to 57.5\%, although the pattern was not strictly monotonic. DeepSeek-V3 fluctuated between 81.3\% and 90.0\% and did not exhibit a consistent decrease across the five levels.

To quantify these patterns, we computed Spearman's rank-order correlation between the number of injected smells and the average and best pass rates. Each correlation was based on the five density-level observations for one application, LLM, and performance metric. GPT-4o exhibited strong negative correlations for average performance in \textit{Snake} ($\rho=-0.97$, exact $p=0.033$) and \textit{Scopa} ($\rho=-0.90$, exact $p=0.083$). Neither correlation remained statistically significant after Bonferroni correction ($\alpha^*=0.0033$).

For DeepSeek-V3, the average pass rates in \textit{Dice} and \textit{Snake} decreased monotonically across the five density levels, resulting in $\rho=-1.00$ in both cases. Given the small number of observations and the boundary value of the coefficient, we used an exact permutation test, which yielded a two-sided $p=0.0167$ for each correlation. These results also did not remain statistically significant after Bonferroni correction. Correlations for \textit{Arkanoid} and \textit{Scopa} were weaker or non-monotonic and were not statistically significant.

For GPT-4o on \textit{Snake}, best-case performance remained constant at 71.4\% across all density levels. Consequently, a meaningful rank correlation could not be computed for this configuration because the outcome exhibited no variation.

\begin{tcolorbox}[colback=gray!5!white, colframe=gray!70!white, title=\textbf{Answer to RQ2}]
\textit{Higher smell density was generally associated with lower average test-suite-based functional correctness, but the magnitude and consistency of this pattern varied across applications and LLMs. None of the correlations remained statistically significant after correction for multiple comparisons.}
\end{tcolorbox}

\subsection{Effect of Smell Categories on LLM Functional Correctness}  

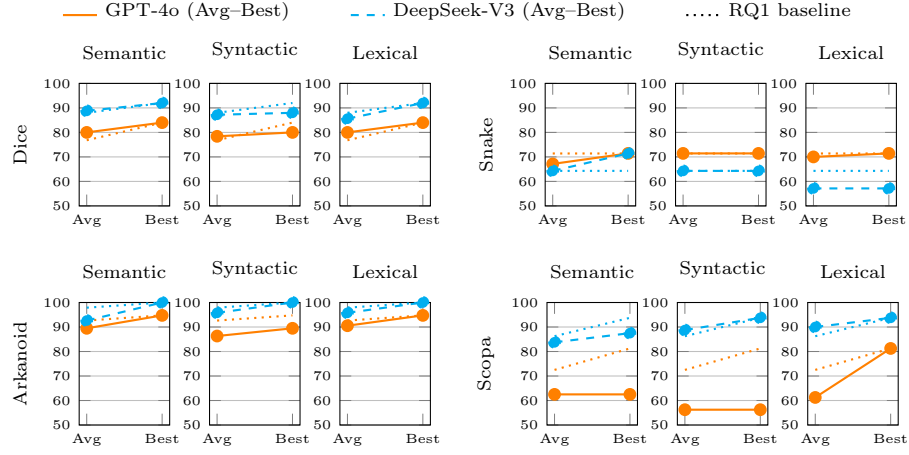
\begin{figure}[t]
\Description{A 2-by-2 grid of game-level plots. Each game contains three subplots for semantic, syntactic, and lexical requirement smell categories. Each subplot shows average and best pass rates for GPT-4o and DeepSeek-V3, with dotted horizontal lines indicating the clean baseline from RQ1.}

\centering
{\scriptsize
\tikz{\draw[orange,thick] (0,0)--(.45,0);} GPT-4o (Avg--Best)
\hspace{0.6cm}
\tikz{\draw[cyan,thick,dashed] (0,0)--(.45,0);} DeepSeek-V3 (Avg--Best)
\hspace{0.6cm}
\tikz{\draw[black,dotted,thick] (0,0)--(.45,0);} RQ1 baseline
}

\vspace{4pt}

\begin{minipage}[t]{0.495\columnwidth}
\centering
\begin{tikzpicture}
\begin{groupplot}[
  group style={group size=3 by 1,horizontal sep=15pt,ylabels at=edge left},
  width=0.46\linewidth,
  height=0.53\linewidth,
  ymin=50,
  ymax=100,
  ytick={50,60,70,80,90,100},
  symbolic x coords={Avg,Best},
  xtick=data,
  tick label style={font=\tiny},
  label style={font=\scriptsize},
  title style={font=\scriptsize},
  every axis plot/.append style={thick},
  ymajorgrids
]

\nextgroupplot[ylabel={Dice},title={Semantic}]
\addplot[color=orange,dotted,thick,forget plot] coordinates {(Avg,76.8) (Best,84.0)};
\addplot[color=cyan,dotted,thick,forget plot] coordinates {(Avg,88.0) (Best,92.0)};
\addplot[color=orange,mark=*] coordinates {(Avg,80.0) (Best,84.0)};
\addplot[color=cyan,dashed,mark=*] coordinates {(Avg,88.8) (Best,92.0)};

\nextgroupplot[title={Syntactic}]
\addplot[color=orange,dotted,thick,forget plot] coordinates {(Avg,76.8) (Best,84.0)};
\addplot[color=cyan,dotted,thick,forget plot] coordinates {(Avg,88.0) (Best,92.0)};
\addplot[color=orange,mark=*] coordinates {(Avg,78.4) (Best,80.0)};
\addplot[color=cyan,dashed,mark=*] coordinates {(Avg,87.2) (Best,88.0)};

\nextgroupplot[title={Lexical}]
\addplot[color=orange,dotted,thick,forget plot] coordinates {(Avg,76.8) (Best,84.0)};
\addplot[color=cyan,dotted,thick,forget plot] coordinates {(Avg,88.0) (Best,92.0)};
\addplot[color=orange,mark=*] coordinates {(Avg,80.0) (Best,84.0)};
\addplot[color=cyan,dashed,mark=*] coordinates {(Avg,85.6) (Best,92.0)};

\end{groupplot}
\end{tikzpicture}
\end{minipage}
\hfill
\begin{minipage}[t]{0.495\columnwidth}
\centering
\begin{tikzpicture}
\begin{groupplot}[
  group style={group size=3 by 1,horizontal sep=15pt,ylabels at=edge left},
  width=0.46\linewidth,
  height=0.53\linewidth,
  ymin=50,
  ymax=100,
  ytick={50,60,70,80,90,100},
  symbolic x coords={Avg,Best},
  xtick=data,
  tick label style={font=\tiny},
  label style={font=\scriptsize},
  title style={font=\scriptsize},
  every axis plot/.append style={thick},
  ymajorgrids
]

\nextgroupplot[ylabel={Snake},title={Semantic}]
\addplot[color=orange,dotted,thick,forget plot] coordinates {(Avg,71.4) (Best,71.4)};
\addplot[color=cyan,dotted,thick,forget plot] coordinates {(Avg,64.3) (Best,64.3)};
\addplot[color=orange,mark=*] coordinates {(Avg,67.14) (Best,71.43)};
\addplot[color=cyan,dashed,mark=*] coordinates {(Avg,64.29) (Best,71.43)};

\nextgroupplot[title={Syntactic}]
\addplot[color=orange,dotted,thick,forget plot] coordinates {(Avg,71.4) (Best,71.4)};
\addplot[color=cyan,dotted,thick,forget plot] coordinates {(Avg,64.3) (Best,64.3)};
\addplot[color=orange,mark=*] coordinates {(Avg,71.43) (Best,71.43)};
\addplot[color=cyan,dashed,mark=*] coordinates {(Avg,64.29) (Best,64.29)};

\nextgroupplot[title={Lexical}]
\addplot[color=orange,dotted,thick,forget plot] coordinates {(Avg,71.4) (Best,71.4)};
\addplot[color=cyan,dotted,thick,forget plot] coordinates {(Avg,64.3) (Best,64.3)};
\addplot[color=orange,mark=*] coordinates {(Avg,70.00) (Best,71.43)};
\addplot[color=cyan,dashed,mark=*] coordinates {(Avg,57.14) (Best,57.14)};

\end{groupplot}
\end{tikzpicture}
\end{minipage}

\vspace{6pt}

\begin{minipage}[t]{0.495\columnwidth}
\centering
\begin{tikzpicture}
\begin{groupplot}[
  group style={group size=3 by 1,horizontal sep=15pt,ylabels at=edge left},
  width=0.46\linewidth,
  height=0.53\linewidth,
  ymin=50,
  ymax=100,
  ytick={50,60,70,80,90,100},
  symbolic x coords={Avg,Best},
  xtick=data,
  tick label style={font=\tiny},
  label style={font=\scriptsize},
  title style={font=\scriptsize},
  every axis plot/.append style={thick},
  ymajorgrids
]

\nextgroupplot[ylabel={Arkanoid},title={Semantic}]
\addplot[color=orange,dotted,thick,forget plot] coordinates {(Avg,92.63) (Best,94.74)};
\addplot[color=cyan,dotted,thick,forget plot] coordinates {(Avg,97.89) (Best,100.0)};
\addplot[color=orange,mark=*] coordinates {(Avg,89.47) (Best,94.74)};
\addplot[color=cyan,dashed,mark=*] coordinates {(Avg,92.63) (Best,100.0)};

\nextgroupplot[title={Syntactic}]
\addplot[color=orange,dotted,thick,forget plot] coordinates {(Avg,92.63) (Best,94.74)};
\addplot[color=cyan,dotted,thick,forget plot] coordinates {(Avg,97.89) (Best,100.0)};
\addplot[color=orange,mark=*] coordinates {(Avg,86.32) (Best,89.47)};
\addplot[color=cyan,dashed,mark=*] coordinates {(Avg,95.79) (Best,100.0)};

\nextgroupplot[title={Lexical}]
\addplot[color=orange,dotted,thick,forget plot] coordinates {(Avg,92.63) (Best,94.74)};
\addplot[color=cyan,dotted,thick,forget plot] coordinates {(Avg,97.89) (Best,100.0)};
\addplot[color=orange,mark=*] coordinates {(Avg,90.53) (Best,94.74)};
\addplot[color=cyan,dashed,mark=*] coordinates {(Avg,95.79) (Best,100.0)};

\end{groupplot}
\end{tikzpicture}
\end{minipage}
\hfill
\begin{minipage}[t]{0.495\columnwidth}
\centering
\begin{tikzpicture}
\begin{groupplot}[
  group style={group size=3 by 1,horizontal sep=15pt,ylabels at=edge left},
  width=0.46\linewidth,
  height=0.53\linewidth,
  ymin=50,
  ymax=100,
  ytick={50,60,70,80,90,100},
  symbolic x coords={Avg,Best},
  xtick=data,
  tick label style={font=\tiny},
  label style={font=\scriptsize},
  title style={font=\scriptsize},
  every axis plot/.append style={thick},
  ymajorgrids
]

\nextgroupplot[ylabel={Scopa},title={Semantic}]
\addplot[color=orange,dotted,thick,forget plot] coordinates {(Avg,72.50) (Best,81.25)};
\addplot[color=cyan,dotted,thick,forget plot] coordinates {(Avg,86.25) (Best,93.75)};
\addplot[color=orange,mark=*] coordinates {(Avg,62.50) (Best,62.50)};
\addplot[color=cyan,dashed,mark=*] coordinates {(Avg,83.75) (Best,87.50)};

\nextgroupplot[title={Syntactic}]
\addplot[color=orange,dotted,thick,forget plot] coordinates {(Avg,72.50) (Best,81.25)};
\addplot[color=cyan,dotted,thick,forget plot] coordinates {(Avg,86.25) (Best,93.75)};
\addplot[color=orange,mark=*] coordinates {(Avg,56.25) (Best,56.25)};
\addplot[color=cyan,dashed,mark=*] coordinates {(Avg,88.75) (Best,93.75)};

\nextgroupplot[title={Lexical}]
\addplot[color=orange,dotted,thick,forget plot] coordinates {(Avg,72.50) (Best,81.25)};
\addplot[color=cyan,dotted,thick,forget plot] coordinates {(Avg,86.25) (Best,93.75)};
\addplot[color=orange,mark=*] coordinates {(Avg,61.25) (Best,81.25)};
\addplot[color=cyan,dashed,mark=*] coordinates {(Avg,90.00) (Best,93.75)};

\end{groupplot}
\end{tikzpicture}
\end{minipage}

\caption{Pass rates by smell category for each game application. Dotted horizontal lines indicate the corresponding non-smelly baseline from RQ1.}
\label{fig:rq3-by-type}
\end{figure}

Figure~\ref{fig:rq3-by-type} compares category-specific performance with the corresponding non-smelly baseline. DeepSeek-V3 achieved higher average performance than GPT-4o across all smell categories for \textit{Dice}, \textit{Arkanoid}, and \textit{Scopa}, whereas GPT-4o achieved higher average performance for \textit{Snake}. However, the differences between smell categories were not consistent across applications or LLMs.

For \textit{Dice}, the category-specific average pass rates remained close to the non-smelly baseline for both LLMs. For \textit{Arkanoid}, all category-specific configurations maintained relatively high pass rates, although GPT-4o showed its largest descriptive decrease under syntactic smells. For \textit{Snake}, GPT-4o showed its largest decrease under semantic smells, while DeepSeek-V3 showed its largest decrease under lexical smells. For \textit{Scopa}, GPT-4o exhibited lower average performance for all three categories, with the largest decrease under syntactic smells. DeepSeek-V3 remained comparatively close to or above its non-smelly average under the syntactic and lexical conditions.

We used chi-square tests of independence to examine associations between smell presence and requirement-level test outcomes. We applied Bonferroni correction separately to two families of comparisons. At the aggregate LLM level, six tests were conducted, corresponding to three smell categories across two LLMs, resulting in a corrected significance threshold of $\alpha^* = 0.05/6 = 0.0083$. No statistically significant aggregate associations were observed, and the corresponding Cramér's V values indicated weak associations. At the application level, 24 tests were conducted, corresponding to four applications, three smell categories, and two LLMs. The resulting corrected threshold was $\alpha^* = 0.05/24 \approx 0.0021$. The smallest uncorrected $p$-value was observed for lexical smells with DeepSeek-V3 in \textit{Snake} ($p = 0.0098$, $V = 0.577$). Although this result indicated a comparatively large effect size, it did not remain statistically significant after correction for multiple comparisons.

\begin{tcolorbox}[colback=gray!5!white, colframe=gray!70!white,
title=\textbf{Answer to RQ3}]
\textit{The descriptive effects of smell categories varied across applications and LLMs, and no category produced a consistently larger decrease in test-suite-based functional correctness. None of the observed associations remained statistically significant after correction for multiple comparisons.}
\end{tcolorbox}

\section{Discussion}
\label{sec:discussion}

Our results show that, under non-smelly conditions, GPT-4o and DeepSeek-V3 generated implementations that passed a majority of the requirement-level tests, particularly for \textit{Arkanoid}. This finding is aligned with prior work on automated traceability, where LLMs also achieved high performance when working with non-smelly requirements~\cite{vogelsang2025impact}. Together, these results suggest that high-quality requirements support successful LLM-assisted SE tasks in controlled settings. However, high baseline performance did not eliminate recurring failures, particularly for requirements involving shared functions, ordered dependencies, or dynamic game states.

\textbf{Requirement Smells Across SE Tasks.} 
Compared with the automated traceability study by Vogelsang \textit{et al.}~\cite{vogelsang2025impact}, our results suggest that requirement smells may have a stronger descriptive impact in code generation tasks. Whereas the traceability study reported only small effects on binary traceability accuracy and no significant effects on LOC tracing F1-score, our study observed larger descriptive decreases in test-suite-based functional correctness for several game–LLM combinations as smell density increased. However, because neither study consistently found statistically significant effects after correction for multiple comparisons, this comparison should be interpreted as exploratory rather than conclusive.

\textbf{Prompt Quality and Task Sensitivity.} 
The results of RQ2 indicate a general tendency for higher smell density to coincide with lower test-suite-based functional correctness in several LLM–application combinations, although the magnitude and consistency of this pattern varied across games. None of the observed correlations remained statistically significant after correction for multiple comparisons. Consequently, these findings should be interpreted as exploratory evidence that requirement smells may influence code generation quality. The descriptive decreases were more evident for \textit{Snake} and \textit{Scopa}. One possible explanation is that these applications involve more interconnected behavior, increasing the opportunity for requirement defects to influence generated implementations. However, task complexity was not experimentally manipulated and therefore this interpretation remains a hypothesis for future work.

\textbf{Smell Categories.} 
Our results indicate that the effects of requirement smell categories varied across applications and LLMs. No smell category consistently produced larger decreases in test-suite-based functional correctness. This suggests that smell sensitivity may emerge from the interaction between requirement quality, application characteristics, and LLM behavior rather than from the smell category alone. Future work should investigate which contextual factors make particular smell types more disruptive in different SE tasks.

\textbf{LLM Evolution and Generalizability.}
LLM capabilities evolve rapidly and newer LLMs may achieve higher code generation performance and be more robust to low-quality inputs, for example by better resolving ambiguity or inferring missing context. Nevertheless, our primary focus is not absolute LLM performance, but how generated-code correctness changes under controlled variations in requirement quality. Whether newer and more capable LLMs exhibit similar sensitivity to requirement smells remains an open empirical question.

\section{Threats to Validity}
\label{sec:threats}

\textbf{Construct validity.} Test-suite-based functional correctness, as measured by requirement-level unit tests, may not capture partial or alternative correct implementations that satisfy the requirements through different implementation strategies. While our use of predefined skeleton code aims to reduce ambiguity in implementation, it may also constrain LLM creativity and bias the results toward predefined structures. Additionally, the selected smell categories and injected variants may only partially represent the broader notion of requirement smells. While the smell injections were designed to resemble realistic defects reported in RE literature and industrial practice, they remain controlled experimental manipulations and may not fully capture the contextual complexity of naturally occurring requirement defects.

\textbf{Internal validity.} Although we systematically injected smells using a controlled prompt structure, interactions between smells and residual LLM stochasticity may still have influenced the results. Requirements ordering may also influence results due to positional biases in LLMs. Additionally, because the selected games are common programming exercises with publicly available implementations on platforms such as GitHub, the evaluated LLMs may have been exposed during training to similar implementations or problem descriptions. Although we cannot determine whether such exposure occurred, it could have influenced baseline performance. Finally, all experiments employed a single prompt template and interaction strategy. Different prompting approaches (\textit{e.g.}, iterative prompting, chain-of-thought prompting, or agent-based workflows) may exhibit different levels of robustness to requirement smells.

\textbf{External validity.} We used four game applications with relatively short and well-scoped functional requirements. While this setup enabled tight control and automated evaluation, it does not reflect the complexity of industrial-scale software systems. Additionally, only two LLMs were evaluated (GPT-4o and DeepSeek-V3), which limits generalizability across the broader landscape of foundation models. Caution should be exercised when applying these findings to other LLMs or domains without additional validation. Furthermore, the experiment used controlled smell injections rather than naturally occurring industrial requirements, which may limit the generalizability of the observed effects to real-world RE settings.

\textbf{Conclusion validity.}
We used Spearman's rank-order correlation and chi-square tests of independence to analyze the relationships between requirement smells and test-suite-based functional correctness. Because the smell-density analysis was based on only five density levels and the smell-category analysis involved relatively small numbers of requirements, statistical power was limited. Consequently, non-significant results should not be interpreted as evidence of the absence of an effect. To reduce the likelihood of false positive findings, we applied exact permutation tests for boundary Spearman correlations and Bonferroni correction separately to the predefined families of comparisons. Finally, comparisons with prior studies on automated traceability should be interpreted cautiously because the studies investigated different SE tasks, evaluation metrics, and statistical analyses.
\section{Concluding Remarks}
\label{sec:concluding-remakrs}

This paper presented a controlled experiment investigating how requirement quality influences the test-suite-based functional correctness of LLM-generated code. Building on prior work on automated traceability between requirements and code~\cite{vogelsang2025impact}, we extended the investigation to a code generation task using four game applications while systematically varying requirement smell density and smell categories.

Regarding \textit{RQ1}, GPT-4o and DeepSeek-V3 achieved high pass rates when prompted with non-smelly requirements, particularly for \textit{Arkanoid}. However, neither LLM consistently produced implementations that satisfied all requirement-level tests, indicating that clean requirements alone do not guarantee complete correctness.

Regarding \textit{RQ2}, higher smell density was generally associated with lower test-suite-based functional correctness across several LLM--application combinations. However, the magnitude and consistency of this pattern varied across games and LLMs, and none of the observed correlations remained statistically significant after correction for multiple comparisons.

Regarding \textit{RQ3}, the effects of smell categories varied across applications and LLMs. No smell category consistently produced larger decreases in test-suite-based functional correctness, and none of the observed associations remained statistically significant after correction for multiple comparisons. These results suggest that the impact of requirement smells may depend on the interaction between requirement quality, task characteristics, and LLM behavior rather than on smell category alone.

Overall, this work contributes empirical evidence on the relationship between requirement quality and LLM-assisted code generation while providing an extension of prior research on requirement smells in AI-assisted SE. Future work should investigate these effects using larger and more realistic applications, additional LLMs, naturally occurring industrial requirements, and other SE tasks to better understand when and how requirement quality influences AI-assisted development.

\bibliographystyle{splncs04}
\bibliography{lipics-v2021-sample-article}

\end{document}